\documentclass{article}
\usepackage{amsfonts}

\newcommand{\bls}[1]{\renewcommand{\baselinestretch}{#1}}
\def\noi{\noindent}

\makeatletter
\renewcommand{\section}{\@startsection{section}{1}{0pt}%
        {-3.5ex plus -1ex minus -.2ex}{2.3ex plus .2ex}%
        {\large\bf\protect\raggedright}}

\renewcommand{\subsection}{\@startsection{subsection}{2}{0pt}%
        {-3ex plus -1ex minus -.2ex}{1.4ex plus .2ex}%
        {\normalsize\bf\protect\raggedright}}

\renewcommand{\thesubsubsection}%
    {\arabic{section}.\arabic{subsection}.\arabic{subsubsection}}

\renewcommand{\@oddhead}{\raisebox{0pt}[\headheight][0pt]{%
   \vbox{\hbox to\textwidth{\rightmark \hfil \rm \thepage \strut}\hrule}}}
\renewcommand{\@evenhead}{\raisebox{0pt}[\headheight][0pt]{%
   \vbox{\hbox to\textwidth{\thepage \hfil \leftmark \strut}\hrule}}}
\makeatother

\newcommand{\Picture}[3]{
	\begin{figure} \centering \unitlength=1mm
	\begin{picture}(125,#1)
		\put(0,0){\line(0,1){#1}}       %frame  % #1 = height (mm)
		\put(0,0){\line(1,0){125}}              % #2 = picture
		\put(125,0){\line(0,1){#1}}             % #3 = caption
		\put(0,#1){\line(1,0){125}}
	\put(0,0){#2}                       \end{picture}
        \caption{\protect\small #3}  \medskip \hrule \end{figure} }
\def\fig{Fig.\,}

\newcommand{\sect}[1]{Sec.\,#1}

\def\nqq{\hspace{-2em}}
\def\nhq{\hspace{-0.5em}}

\def\cm{\hspace{1cm}}
\def\inch{\hspace{1in}}

\newcommand{\sequ}[1]{\setcounter{equation}{#1}}
\def\eqdef{\stackrel{\rm def}=}

\def\eq{Eq.\,}

\def\beq{\begin{equation}}
\def\eeq{\end{equation}}
\def\bear{\begin{eqnarray}}
\def\al{&\nhq}
\def\lal{&&\nqq {}}               % left alignment
\def\bearr{\begin{eqnarray} \lal}
\def\ear{\end{eqnarray}}
\def\earn{\nonumber \end{eqnarray}}
\def\dst{\displaystyle}
\def\tst{\textstyle}

\def\nn{\nonumber\\ {}}
\def\nnv{\nonumber\\[5pt] {}}
\def\nnn{\nonumber\\ \lal }

\def\yy{\\[5pt]}

\def\eql{\al =\al}

\def\e{{\,\rm e}}
\def\d{\partial}

\def\sign{\mathop{\rm sign}\nolimits}

\def\const{{\rm const}}
\def\Half{{\dst\frac{1}{2}}}
\def\half{{\tst\frac{1}{2}}}
\def\then{\quad \Longrightarrow\quad }
\def\DAL{\mathop{\raisebox{3.5pt}{\large\fbox{}}}\nolimits}
\newcommand{\vars}[1]{\left\{\begin{array}{ll}#1\end{array}\right.}

\newcommand{\limas}[1]{\mathop{\lim}\limits_{#1}}
\newcommand{\Theorem}[2]{\medskip\noi {\bf #1. \ }{\sl #2}\medskip}

\def\mn{{\mu\nu}}
\def\ep{\epsilon}

\def\og{{\overline g}{}}
\def\R{{\mathbb R}}
\def\df{\delta\varphi}
\def\da{\delta\alpha}
\def\db{\delta\beta}
\def\dg{\delta\gamma}
\def\pn{perturbation}
\def\pns{perturbations}

\def\sph{spherically symmetric\ }
\def\wh{wormhole}
\def\whs{wormholes}
\def\bh{black hole}
\def\bhs{black holes}
\def\TH{\mbox{$T_{\rm H}$}}

\def\RN{Reissner-Nordstr\"om}
\def\umx{u_{\max}}

\def\shhsq{\sinh^2[h(u+u_1)]}

\bls{1.04}

\begin{document}
\thispagestyle{empty}
\rightline{\bf gr-qc/9903028}
\bigskip

\begin{center}
{\LARGE
 ELECTRICALLY CHARGED COLD BLACK HOLES \yy IN SCALAR-TENSOR THEORIES}
\bigskip

{\bf K.A. Bronnikov%
\footnote{e-mail: kb@cce.ufes.br;
permanent address: Centre for Gravitation and Fundamental Metrology,
VNIIMS, 3-1 M. Ulyanovoy St., Moscow 117313, Russia, e-mail:
kb@rgs.mccme.ru}
C.P. Constantinidis\footnote{e-mail: clisthen@cce.ufes.br},
R.L. Evangelista\footnote{e-mail: rleone@cce.ufes.br} and
J.C. Fabris\footnote{e-mail: fabris@cce.ufes.br}} \medskip

{\it Departamento de F\'{\i}sica, Universidade Federal do Esp\'{\i}rito
Santo, Vit\'oria, Esp\'{\i}rito Santo, Brazil}
\medskip

\end{center}

\begin{abstract}
We study the possible existence of charged black holes
in the Bergmann-Wagoner class of scalar-tensor theories (STT) of gravity in
four dimensions.  The existence of \bhs\ is shown for anomalous versions of
these theories, with a negative kinetic term in the Lagrangian.  The Hawking
temperature \TH\ of these holes is zero, while the horizon area is (in most
cases) infinite. As a special case, the Brans-Dicke theory is studied in
more detail, and two kinds of infinite-area \bhs\ are revealed, with finite
and infinite proper time needed for an infalling particle to reach the
horizon; among them, analyticity properties select a discrete subfamily of
solutions, parametrized by two integers, which admit an extension beyond
the horizon. The causal structure and stability of these solutions with
respect to small radial \pns\ is discussed.  As a by-product, the stability
properties of all \sph electrovacuum STT solutions are outlined.
\end{abstract}

%%%%%%%%%%%%%%%%%%%%%%%%%%%%%%%%%%%%%%%%%%%%%%%%%%%%%%%%%%%%%%%%%%%%%%%%%

\section{Introduction}  % S1

This study was to a certain extent stimulated by a controversy in the
recent literature:  the paper by Campanelli and Lousto \cite{lousto}
asserts that in the well-known family of static, \sph vacuum solutions of
the Brans-Dicke (BD) theory there exists a subfamily which possesses all
properties of \bh\ (BH) solutions, but (i) these solutions exist only for
negative values of the coupling constant $\omega$ and (ii) the horizons
have an infinite area. These authors argue that large negative $\omega$
are compatible with modern observations and that such \bhs\ may be of
astrophysical relevance. On the other hand, H. Kim and Y. Kim
\cite{kim}, agreeing that there are non-Schwarzschild \bhs\ in the
Brans-Dicke theory, claim that such \bhs\ have unacceptable thermodynamical
and geometric properties and are therefore physically irrelevant;
meanwhile, they ascribe such solutions to positive values of $\omega$.

The aim of this study is not only to make the situation clear, but a bit
wider: to reveal possible vacuum and electrically charged \bh\
solutions among static, spherically symmetric solutions of the general
(Bergmann-Wagoner) class of scalar-tensor theories (STT) of gravity, which
may be described in terms of the coupling function $\omega(\phi)$; the
BD theory ($\omega=\const$) will be used as the most well-known
example. One of the reasons for such an approach is that, by modern views,
it is rather probable that $\omega$ could have been sufficienly small and
could appreciably affect the physical processes in the early Universe, but
by now became large, making the theory very close to general relativity
(GR) in observational predictions \cite{spectra}.

    We show, in the framework of the general STT, that nontrivial BH
    solutions can exist for the coupling function $\omega(\phi)+3/2 <0$, and
    that only in exceptional cases these BHs have a finite horizon area.

    The case of the BD theory is studied in more detail.
    Various types of geometry are indicated, including
    BHs, wormholes and ``hornlike'' structures, all of them existing in the
    anomalous case $\omega<-3/2$.
    All nontrivial (with the scalar field $\phi \ne \const$) BHs have
    infinite horizon areas%
\footnote
{For brevity, we call BHs with infinite horizon areas type B BHs \cite{we},
to distingish them from the conventional ones,
with finite horizon areas, to be called type A.}
    and zero Hawking temperature (``cold BHs''), thus
    confirming the conclusions of \cite{lousto}.  These BHs in turn split
    into two subclasses: B1, where horizons are attained by infalling
    particles in a finite proper time $\tau$, and B2, for which $\tau$
    is infinite.

    The static region of a type B2 BH is geodesically complete since its
    horizon is infinitely remote and actually forms a second spatial
    asymptotic. For type B1 BHs the global picture is more complex and is
    discussed in some detail. It turns out that the horizon
    is generically singular due to violation of analyticity, despite the
    vanishing curvature invariants. Only a discrete set of B1-solutions,
    parametrized by two integers $m$ and $n$, admits a Kruskal-like
    extension, and, depending on their parity, four different global
    structures are distinguished. Two of them, where $m-n$ is even, are
    globally regular, in two others the region beyond the horizon contains a
    spacelike or null singularity.

    All BHs under consideration turn out to be stable under small radial
    perturbations.

    Since the vacuum case has been described in detail in our previous papers
    \cite{kg1,kg2}, we concentrate here on the solutions with nonzero
    electric charge, only refering to vacuum configurations as a
    limiting case of the charged ones.

The paper is organized as follows.

In \sect 2 we discuss the criteria used to single out \bh\ metrics among
other static, \sph metrics. In \sect 3 we present the (well-known)
electrovacuum solution of the general STT and its vacuum counterpart.
\sect 4 is devoted to a search for possible \bhs\ in
the framework of the general STT. In \sect 5 we outline the properties of
various electrovacuum BD\ solutions, paying special attention to \bh\ ones.
The properties of electrovacuum BD\ \bhs\ are discussed
in \sect 6. In \sect 7 we investigate the stability of the above STT
solutions under radial perturbations and, in particular, show that the BH
and wormhole solutions are stable. \sect 8 contains some concluding remarks.

In the Appendix it is explicitly shown that, in an arbitrary
static, \sph space-time, an infinite Hawking temperature can occur only at a
curvature singularity, and the regularity requirement implies the
invisibility of a horizon for an observer at rest.

%%%%%%%%%%%%%%%%%
\section{Criteria for black hole selection}  % S2
%%%%%%%%%%%%%%%%%

We will deal with static, \sph space-times, whose metric in a
general may be written as
\beq                                                         \label{m1}
   ds^2 = g_\mn dx^{\mu}dx^{\nu}
   = \e^{2\gamma}dt^2 - \e^{2\alpha}du^2 - \e^{2\beta}d\Omega^2
\eeq
where $\gamma$, $\alpha$ and $\beta$ are functions of $u$ only
and $d\Omega^2= d\theta^2 + \sin^2\theta\; d\phi^2$.

Black hole (BH) solutions with the metric (\ref{m1}) are conventionally
singled out among other solutions
by the following criteria: at some surface $u=u^*$ (horizon)
\begin{description}
\item [C1.]  $\e^\gamma$ $\to  0$ (the timelike Killing vector becomes
              null).
\item [C2.]  $\e^\beta$ is finite (finite horizon area).
\item [C3.]  The integral $t^* = \int \e^{\alpha - \gamma} \to \infty$
             as $u \to  u^*$ (invisibility of the horizon for an
	     observer at rest).
\end{description}

      The evident requirement that a horizon must be a regular
      surface (otherwise we deal with a singularity rather than a horizon)
      creates two more criteria:
\begin{description}
\item [C4.]  The Hawking temperature $\TH$ is finite;
\item [C5.]  The Kretschmann scalar $K$ is finite at $u=u^*$.
\end{description}
%
% In addition to these criteria, we will use in \sect 5 the positive mass
% requirement, evident from physical considerations.

As shown in the Appendix, the conditions C3 and C4 are necessary but not
sufficient for regularity of a candidate horizon (a surface where
$\e^{\gamma}=0$), so they will be used as convenient selection tools.
As for C5, the scalar $K$, due to its structure, is the most reliable probe
for space-time regularity.

The condition C2 is apparently less essential than the others.
In principle, C2 can be cancelled, leading to
a generalized notion of a BH, that with a horizon having an
infinite area, as described in \cite{lousto}.
We will call the BHs satisfying all the criteria C1--C5 type A \bhs,
and those with an infinite horizon --- type B \bhs.

We shall see that in the general scalar-tensor theory (STT) most of
nontrivial (non-Schwarzschild and non-\RN) black holes
are type B. In particular, in the BD\ theory all of them are type
B, while all configurations satisfying C1--C3 turn out to be singular.

%%%%%%%%%%%%%%%%%%%%
\section{The generalized \RN\ solution}  % S3

A general Lagrangian describing the interaction between gravity and
a scalar field in the presence of an electromagnetic field in four dimensions
can be written as
\beq                                               \label{L1}
L = \sqrt{-g}\biggr(f(\phi)R
         + \frac{\omega(\phi)}{\phi}\phi_{;\mu}\phi^{;\mu}
		- F_{\mu\nu}F^{\mu\nu}\biggl)
\eeq
where $f(\phi)$ and $\omega(\phi)$ are, in principle, arbitrary
functions of the scalar field $\phi$ (the so-called Bergmann-Wagoner
class of STT). Reparametrization of $\phi$ makes it possible to leave
only one arbitrary function; a conventional choice
is such that $f=\phi$ and $\omega(\phi)$ remains arbitrary.
We will use it here as well.

The formulation (\ref{L1}) (the so-called {\it Jordan conformal
frame\/}) is commonly considered to be fundamental since just in this
frame the matter energy-momentum tensor $T^\mu_\nu$ obeys the
conventional conservation law $\nabla_{\alpha}T^\alpha_\mu =0$, leading
to the usual equations of motion (the so-called atomic system of
measurements). In particular, free particles move along geodesics of
the Jordan-frame metric. Therefore, in what follows {\it we discuss
the geometry and causal structure of the solutions in the Jordan
frame.}

The field equations are easier to deal with in
the {\it Einstein conformal frame\/}, where the transformed scalar
field $\varphi$ is minimally coupled to gravity. Namely, the
conformal mapping $g_{\mu\nu} = \phi^{-1}\og_{\mu\nu}$ transforms \eq
(\ref{L1}) (up to a total divergence) to
\bearr                                                  \label{L2}
     L = \sqrt{-\og}\biggr(\overline{R} + \ep
         \og^{\alpha\beta}\varphi_{;\alpha}\varphi_{;\beta}
		- F_{\mu\nu}F^{\mu\nu}\biggl),           \\  \lal
             \ep = \sign (\omega + 3/2),                \label{eps}
	\cm
  \frac{d\varphi}{d\phi} = \biggl|\frac{\omega + 3/2}{\phi^2}\biggr|^{1/2}
\ear
where bars mark quantities defined in the Einstein frame; the field
$F_{\mn}$ is not transformed and the indices in (\ref{L2}) are raised
using $\og^{\mn}$. The value $\ep=+1$ corresponds to normal STT, with
positive scalar field energy density in the Einstein frame; the choice
$\ep=-1$ is anomalous. When $\phi = \const$, the theory reduces to GR.

With the aid of the above transformation, the following form of the
exact static, \sph solution to the field equations
due to (\ref{L1}), containing a nonzero electric charge $q$, has
been obtained \cite{73} (the notations are here slightly changed):
\bear                                                            \label{s2}
     ds^2 \eql g_{\mn} dx^{\mu} dx^{\nu}
             = \phi^{-1}\og_{\mn} dx^{\mu} dx^{\nu}     \nn
	     \eql
     \frac{1}{\phi}\biggl\{ \frac{1}{q^2\,s^2(h,u+u_1)}dt^2 -
     				 \frac{q^2\,s^2(h,u+u_1)}{s^2(k,u)}
          \biggr[\frac{du^2}{s^2(k,u)} + d\Omega^2\biggl]\biggr\}, \yy
     F_{\mn} \eql                                                 \label{F}
	     (\delta_{\mu 0}\delta_{\nu 1}
		-\delta_{\nu 0}\delta_{\mu 1})\,q \e^{\alpha+\gamma-2\beta}
	     = (\delta_{\mu 0}\delta_{\nu 1}
	    -\delta_{\nu 0}\delta_{\mu 1})\,\frac{1}{q\,s^2(h,u+u_1)};\yy
     \varphi\eql Cu, \cm                                         \label{phi}
	\frac{\omega+ 3/2}{\phi^2} \biggl(\frac{d\phi}{du}\biggr)^2
		 =\ep C^2
\ear
     where the constant $C$ has the meaning of a scalar charge.
     The integration constants $C,\ k,\ h$ are related by
\beq
     2k^2\sign k = \ep C^2 + 2h^2\sign h .            \label{r2}
\eeq
	The function $s(k,u)$ is defined as follows:
\beq                                                     \label{f2}
	s(k,u) = \vars     {
                    k^{-1}\sinh ku,  \ & k > 0 \\
                                 u,  \ & k = 0 \\
                    k^{-1}\sin ku,   \ & k < 0.  }
\eeq

Here $u$ is a convenient radial variable (it is a harmonic coordinate
in the Einstein frame). The range of $u$ is $0 < u < \umx$, where $u=0$
corresponds to spatial infinity, while $\umx$ may be finite or infinite
depending on the constants $k,$ $h,$ $u_1$ and the behaviour of
$\phi(\varphi)$ described in (\ref{phi}).

Without losing generality we can normalize $\phi$ to unity at
spatial infinity ($u=0$), so that $\phi(0)=1$, and require $g_{00}(0)
=1$.  The integration constant $u_1$ then satisfies the condition
\beq
	s^2(h,\ u_1) = 1/q^2                              \label{u1}
\eeq
(preserving some discrete arbitrariness of $u_1$).
We thus have three essential integration constants:
$k$ or $h$ and	the charges $q$ and $C$. An expression for the mass $M$
of the configuration is obtained by comparing the asymptotic of
(\ref{s2}) with the Schwarzschild metric and depends on the asymptotic
behaviour of $\omega(\phi)$:
\beq
	GM = \frac{\phi'}{2\phi}\biggr|_{u=0}                   \label{GM}
		+ \frac{s'(h,u+u_1)}{s(h,u+u_1)}\biggr|_{u=0}
	   = \frac{\pm C}{\sqrt{|\omega(1)+3/2|}}
		\pm \sqrt{q^2 + h^2 \sign h}
\eeq
where $G$ is Newton's gravitational constant. The first ``$\pm$'' sign
reflects the arbitrariness in the sign of $C$ while the second one
depends on the choice of $u_1$ among the variants admitted by
(\ref{u1}).

The \RN\ solution of GR is a special case obtained
herefrom by putting $\phi\equiv 1$, whence $C=0$ and $\varphi\equiv 0$.
Then from (\ref{r2}) it follows $h=k$, and the familiar form of the
\RN\ metric is recovered after the coordinate transformation
\beq
	r = \frac{|q|\,s(k,u+u_1)}{s(k,u)} \then
	\e^{2ku} = \frac{r+k-GM}{r-k-GM}.                       \label{RN}
\eeq

To obtain another limiting case $q=0$ (the scalar-vacuum solution), one
should consider the limit $q\to 0$ preserving the boundary condition
(\ref{u1}). This is only possible for $h \geq 0$ and $u_1\to\infty$.
The resulting metric is
\beq
     ds^2 = \frac{1}{\phi}\biggl\{
               \e^{-2hu}dt^2 - \frac{\e^{2hu}}{s^2(k,u)}
     \biggr[\frac{du^2}{s^2(k,u)} + d\Omega^2\biggl]\biggr\}.   \label{vac}
\eeq
     The scalar field is determined, as before, from (\ref{phi}), and the
     integration constants are related by
\beq                                                      \label{r1}
            2k^2\sign k = 2h^2 + \ep C^2
\eeq
     This is just the scalar-vacuum solution in the form obtained in
     \cite{73}, which has been studied in detail in \cite{kg1, kg2}.

     It should be noted that in (\ref{vac}), (\ref{r1}) the constant $h$ can
     have any sign but in (\ref{GM}) in the vacuum case the second term
     is just $+h$.

%%%%%%%%
\section{Possible black holes in the general scalar-tensor theory} % S4
%%%%%%%%

     Let us analyze the possible existence of nontrivial (i.e.
     non-Schwarzschild and non-\RN)
     BHs in the general STT, i.e. with variable
     $\omega=\omega(\phi)$, using Criteria C1--C5. We assume $C\ne 0$.

     Recalling that the range of $u$ is $0 <u <\umx$, we can assume
     in our search for \bhs\ that $\umx$ is specified by the behaviour of
     $s(k,u)$ and $s(h,u+u_1)$. The opportunity of
     $\phi\to 0$ or $\phi\to\infty$ at some $u=u_0$ inside this range
     must be rejected since in this case, e.g., Criterion C3 is necessarily
     violated at such $u_0$; moreover, when $\phi\to\infty$, it is a
     singular centre, and when $\phi\to 0$, we have $g_{tt}\to\infty$.

     The solutions belonging to the family (\ref{s2})--(\ref{phi}) may be
     then classified as follows:
\bear
	[1+] &&  \ep=+1,\ \ k >h >0 ;     \nn
	[2+] &&  \ep=+1,\ \ k> h=0 ;      \nn
	[3+] &&  \ep=+1,\ \ h<0 ;         \nn
	[1-] &&  \ep=-1,\ \ h > k >0 ;    \nn
	[2-] &&  \ep=-1,\ \ h > k =0 ;    \nn
	[3-] &&  \ep=-1,\ \ h \geq 0,\ k <0 ; \nn
	[4-] &&  \ep=-1,\ \ 0 > h > k .                   \label{class}
\ear

     For the vacuum solution (\ref{vac}), (\ref{phi})
     there exist only four classes:
\bear
	[1+] &&  \ep=+1,\ \ k >0;    \nn
	[1-] &&  \ep=-1,\ \ k >0;    \nn
	[2-] &&  \ep=-1,\ \ k =0;    \nn
	[3-] &&  \ep=-1,\ \ k <0;                	\label{class-vac}
\ear
     here $h \in \R$.

     One can verify that C3 (whose formulation does not depend on
     $\phi(u)$ and hence on the choice of $\omega(\phi)$) is violated for
     all solutions with $\ep=+1$.
     From the viewpoint of Criteria C1--C3, for both vacuum \cite{k1}
     and electrovacuum solutions, there are four opportunities of BH
     existence:

\begin{description}
\item[{$[1-]$}]: $k>0$,  $u^* = \infty$;

\item[{$[1+]$}]:
     $u = \infty$ is a regular sphere and a horizon may
     be found beyond it by proper continuation (example: a BH with a
     conformal scalar field);

\item[{$[2-]$}]: $k=0$, $u^* = \infty$;

\item[{$[3-],\ [4-]$}]: $k<0$, $u^* = \pi/|k|$.
\end{description}

     Let us consider, for $q\ne 0$, each case separately, except the
     second one, since it is hard to handle in a general form due to the
     continuation. We will try first to apply the requirements C1--C3 and
     after that C4 and C5. One can notice that in all cases to be considered
     the theory is anomalous%
\footnote{As shown in \cite{k1}, the cases when $\omega>-3/2$ and the
     sphere $u=\infty$ is regular and admits an extension of the static
     coordinate chart, are very rare, and even when it is the case, such
     configurations turn out to be unstable due to blowing-up of the
     effective gravitational coupling \cite{78}.}.

\medskip\noi
{\bf [1\,--]}.
     In (\ref{s2}) both $s(.,.)$ are hyperbolic sines and $u_1>0$.
     Criteria C1--C3 are satisfied when, as $u\to\infty$,
\beq                                                        \label{4.2}
     \phi \sim \e^{2(h-k)u}, \cm \e^{2\gamma} \sim \e^{(2k-4h)u}.
\eeq
     The Hawking temperature and the term $K_1$ in the Kretschmann scalar
     (see the Appendix) behave as follows:
\beq                                                        \label{4.3}
     \TH = \lim_{u\to\infty} \e^{2(k-h)u} =0; \cm
     K_1 \sim \e^{2ku} \to \infty.
\eeq
\medskip\noi
{\bf [2\,--]}.
     Criteria C1--C3 are satisfied when, as $u\to\infty$,
\beq                                                        \label{4.4}
    \phi \sim \e^{2hu}/u^2, \cm \e^{2\gamma} \sim u^2\e^{-4hu}.
\eeq
     Similarly to (\ref{4.3}), we obtain
\beq                                                        \label{4.5}
     \TH = \lim_{u\to\infty} u^2\e^{-2hu} =0; \cm
     K_1 \sim 2hu^2 \to \infty.
\eeq
\medskip\noi
{\bf [3\,--],\ [4\,--]}.
     A possible horizon is at $u^* = \pi/|k|$, and in its vicinity
     all functions depend on $\Delta u \equiv |u - u^*|$.
     A calculation shows that all the requirements C1--C5 can be satisfied:
\bearr
	\phi \sim (\Delta u)^{-2}, \cm \e^{2\gamma} \sim (\Delta u)^2, \nnn
	\TH = \limas{u\to u^*} \Delta u =0; \cm K_1 < \infty.   \nnn
	\frac{\phi_u}{\phi} \sim \frac{1}{\Delta u} \to  \infty
	     \then  \omega + \frac{3}{2} \to  -0  .           \label{4.6}
\ear
     The behaviour of $g_{00}$ and $g_{11}$ near the horizon is in
     this case similar to that in the extreme \RN\ solution.  This is the
     only case when a type A BH can appear among the STT solutions under
     consideration. As follows from (\ref{4.6}), this opportunity can be
     realized only in a theory with $\omega(\phi) \ne \const$, so the BD\
     theory is excluded.

\medskip
The qualitative picture for $q=0$ (vacuum) \cite{kg1,kg2} is reproduced
if we exclude the class $[4-]$.

We notice that in all these solutions the Hawking temperature calculated for
the assumed horizons is zero. Meanwhile, $\TH < \infty $ is only necessary
but not sufficient for regularity. In all cases with $k\geq 0$, the
Kretschmann scalar tends to infinity, so that the surfaces of finite area,
satisfying the conventional criteria C1--C3 of an event horizon, turn out to
be singular. Therefore {\sl only type B \bhs\ can exist for \ $k\geq 0$}.
This will be demonstrated explicitly for the BD theory, that is, in cases
$[1-]$ and $[2-]$.  As for $k<0$, type A \bhs\ are possible, but such
examples are yet to be found and must be sought for in theories other than
BD.

%%%%%%%%%
\section{Electrovacuum and vacuum Brans-Dicke solutions}     % S5
%%%%%%%%%

Consider the solution (\ref{s2})--(\ref{phi}) and put $\omega =
\const$. Then we can explicitly write
\beq                                                        \label{phi-BD}
    \phi = e^{-2\sigma u}, \cm 2\sigma = C/\sqrt{|\omega + 3/2|}.
\eeq
Let us briefly characterize the properties of the solution using the
classification (\ref{class}) and single out possible BH solutions,
which, as we already know, can be found in the classes $[1-]$ and $[2-]$.

\subsection*{${\bf [1+]:}\ k>h>0$}

     Suppose $u_1 > 0$, then $\umx = \infty$. A positive mass (\ref{GM}) is
     obtained when $h>\sigma$, and then, as $u\to \infty$,
\beq
	\e^{\gamma} \sim \e^{(\sigma-h)u}\to 0, \cm
	\e^{\beta}  \sim \e^{(\sigma+h-k)u}.              \label{6.1}
\eeq
     One can easily verify that $\TH=\infty$. So we have at $u=\infty$ an
     attracting (due to $\e^{\gamma}\to 0$) singularity having a zero,
     finite or infinite area depending on \ $\sign (\sigma+h-k)$.

\subsection*{${\bf [2+]:}\ k > h = 0$}

     For $u_1>0$ we have again a violation of the requirements C3 and C4,
     hence a singularity at $u=\umx=\infty$. In its neighbourhood the metric
     functions behave as follows:
\beq
     \e^{\gamma} \sim \e^{\sigma u}/(u+u_1), \cm
     \e^{\beta} \sim (u+u_1) \e^{(\sigma-k)u}.            \label{6.2}
\eeq
     The singularity is attracting for $\sigma < 0$ and repelling for
     $\sigma > 0$; it has a zero area (i.e. it is a centre) if $\sigma<k$
     and an infinite area if $\sigma \geq k$. The requirement $M>0$ leads to
     just $\sigma < 1/u_1$, which does not forbid any of the qualitatively
     different variants of behaviour.

\subsection*{${\bf [3+]:}\ h < 0$}

     In this case $k$ can have either sign but the qualitative behaviour of
     the solution is in all cases governed by the function
     $\sin [|h|(u+u_1)]$ and $\umx$ is its smallest positive zero.
     The value $u=\umx$ corresponds to a \RN{}-like naked repelling central
     singulariry: $\e^{\gamma}\to \infty$, $\e^{\beta}\to 0$.

     The solutions $[1\pm]$ and $[2\pm]$ with $u_1 < 0$ have $\umx=-u_1$ and
     behave qualitatively in the same way as $[3+]$.
     Therefore in what follows we will omit the opportunity $u_1 < 0$
     when treating the classes $[1\pm]$ and $[2\pm]$.

\subsection*{${\bf [1-]:}\ h>k>0$}

     Now $\umx = \infty$ and, as $u\to \infty$, the metric functions
     and the term $K_1$ of the Kretschmann scalar behave like
\beq
     \e^{\gamma} \sim \e^{(\sigma-h)u};\cm
     \e^{\beta}  \sim \e^{(\sigma+h-k)u};\cm
     K_1 \sim  \e^{2(2k-h-\sigma)u},                     \label{6.3}
\eeq
     so that the regularity condition is $h+\sigma \geq 2k$.
     On the other hand, the condition $\e^\gamma\to 0$ (criterion
     C1) implies $h>\sigma$. Combined, these two conditions give the
     following allowed range of $\sigma$:
\beq
     h > \sigma \geq 2k-h.                               \label{6.4}
\eeq
     As follows from (\ref{6.3}), in this range the area of the surface
     $u=\infty$ is infinite, and, as is easily verified, all the criteria
     C1, C3, C4, C5 are satisfied, so that we have a type B BH with zero
     Hawking temperature \TH.

     Outside the allowed range (\ref{6.4}), when $\sigma \geq h$, there is a
     nonsingular wormhole-like structure where $\e^\beta\to \infty$ as
     $u \to \infty$, while $\e^\gamma\to\infty$ if $\sigma>h$, so that the
     second spatial infinity is repulsive, and $\gamma$ tends to a finite
     limit giving a usual spatial infinity if $\sigma=h$.

     For $\sigma < 2k - h$, the surface $u=\infty$ is singular, attractive
     ($\e^\gamma\to 0)$, and can have a zero, finite or infinite area,
     depending on \ $\sign (\sigma +h - k)$.

\subsection*{${\bf [2-]:}\ h>k=0$}

     Again $\umx=\infty$ and instead of (\ref{6.3}) we have
\beq
     \e^{\gamma} \sim \e^{(\sigma-h)u};\cm
     \e^\beta  \sim \frac{1}{u}\e^{(\sigma+h)u};\cm
     K_1 \sim  u^4 \e^{-2(\sigma+h)u}.                   \label{6.5}
\eeq
     The resulting regularity condition $\sigma+h >0$, combined with
     the requirement $h>\sigma$ that follows from BH criterion C1, yield the
     allowed range of $\sigma$ in the form
\beq
	h > \sigma \geq -h                                \label{6.6}
\eeq
     and, in full similarity to class $[1-]$ with (\ref{6.4}), we have a
     cold (\TH=0) type B \bh.

     Outside the allowed range (\ref{6.6}), for $\sigma\geq h$ we find a
     wormhole-type regular structure similar to the one in the case $[1-]$,
     $\sigma \geq 2h-k$, while for $\sigma < -h $ we find an attracting
     singular centre.

\subsection*{${\bf [3-]:}\ h \geq 0,\ k<0$}

     In this case the solution behaviour is entirely governed by the
     function $\sin |k|u$ and $\umx=\pi/|k|$, while other functions entering
     into the solution are finite and smooth over the whole range of $u$.
     The two asymptotics, $u=0$ and $u=\umx$, are both flat and are
     connected by a regular bridge, so that we are dealing with a
     static, traversable wormhole.

\subsection*{${\bf [4-]:}\ k < h < 0$}

     The qualitative nature of the metric is unaffected by the factor
     $1/\phi=\e^{2\sigma u}$ and is determined by the interplay of
     the two sines: $\sin (|k|u)$ and $\sin [|h|(u+u_1)]$.
     Namely, depending on the positions of their zeros, three cases are
     possible, as shown in Fig.\,1:

\Picture{50}
{\inch
\unitlength=.66mm
%\unitlength=1mm
\special{em:linewidth 0.4pt}
\linethickness{0.4pt}
\begin{picture}(131.00,85.00)(24,70)
\put(29.00,80.00){\vector(1,0){101.00}}
\put(65.00,75.00){\vector(0,1){66.00}}
\put(106.00,75.00){\line(0,1){56.00}}
\put(131.00,85.00){\makebox(0,0)[cc]{$u$}}
\put(62.00,76.00){\makebox(0,0)[cc]{0}}
\put(108.00,76.00){\makebox(0,0)[lc]{$\pi/|k|$}}
\bezier{544}(65.00,80.00)(86.00,145.00)(106.00,80.00)
\bezier{580}(41.00,80.00)(74.00,145.00)(106.00,80.00)
\bezier{580}(29.00,80.00)(62.00,145.00)(94.00,80.00)
\bezier{580}(53.00,80.00)(86.00,145.00)(118.00,80.00)
\put(118.00,101.00){\makebox(0,0)[lb]{$\sin (|k|u)$}}
\put(102.00,93.00){\line(5,2){14.00}}
\put(48.00,123.00){\makebox(0,0)[lb]{[a]}}
\put(72.00,123.00){\makebox(0,0)[lb]{[c]}}
\put(98.00,123.00){\makebox(0,0)[rb]{[b]}}
\put(49.00,120.00){\line(1,-4){2.67}}
\put(73.00,120.00){\line(2,-5){3.33}}
\put(96.00,120.00){\line(1,-4){3.67}}
\end{picture}
}
%\caption
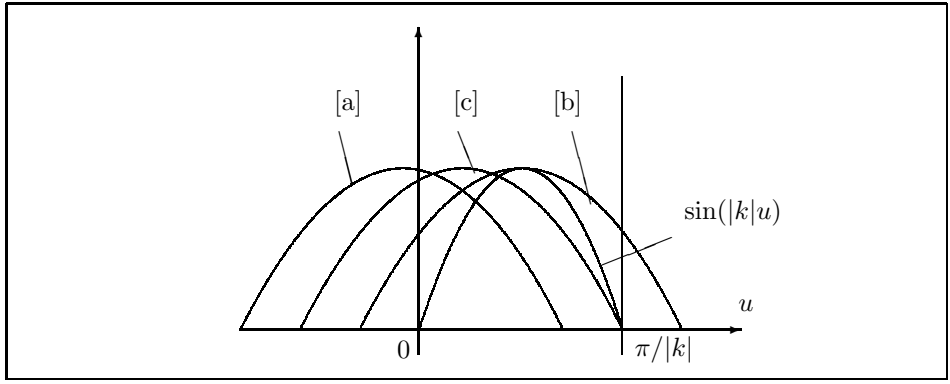
	{Solutions $[4-]$: different positions (a,b,c) of the
	curve $\sin[|h|(u+u_1)]$ with respect to $\sin(|k|u)$ determine
	different behaviours of the solution.}
%\end{figure}

\medskip\noi
     [4--a]: $\umx = \pi/|h|-u_1$ (without loss of generality): the
     solution behaves as that of class $[3+]$.

\medskip\noi
     [4--b]: $\umx = \pi/|k|$: a behaviour like that of class $[3-]$.

\medskip\noi
     [4--c]: $\umx = \pi/|h|-u_1 = \pi/|k|$. As $u\to\umx$,
     $\e^{\gamma}\to\infty$, while $\e^{\beta}$ and the Kretschmann scalar
     tend to finite limits. So we obtain a singularity-free hornlike
     structure (like the ones obtained in some solutions of dilaton gravity
     \cite{horn}), where the infinitely remote
     (since $l=\int \e^{\alpha}du$ diverges) ``end of the horn'', whose
     radius $\e^{\beta}$ is asymptotically constant, repels test particles.

\medskip
     In the vacuum case (\ref{vac}), (\ref{phi-BD}) we are left with the
     classes (\ref{class-vac}), for which the
     estimates (\ref{6.1}), (\ref{6.3})--(\ref{6.6}) remain valid.

     As has been expected, there are type B \bhs\ among the
     solutions of classes $[1-]$ and $[2-]$. However, the family of charged
     BD\ solutions is richer in variants of behaviour as compared with the
     vacuum family: in addition to analogues of vacuum structures, we now
     find repelling \RN{}-like singularities and, in the special case
     [4--c], a nonsingular hornlike structure.

%%%%%%%%%
\section{Charged and neutral Brans-Dicke \bhs}     % S5
%%%%%%%%%

\subsection{Preliminaries}

     We will here study in some detail the cases $[1-]$ and $[2-]$
     of the BD solution of the previous section, when one can indeed find
     some \bh\ type configurations.

     In the case $[1-]$ the metric has the form
\bear                                                    \label{ds1-}
     ds^2 = \e^{2\sigma u}
     \biggl\{ \frac{h^2\,dt^2}{q^2\,\sinh^2[h(u+u_1)]} -
     			\frac{q^2 k^2\,\sinh^2[h(u+u_1)]}
				{h^2 \sinh^2 ku}
          \biggr[\frac{k^2\,du^2}{\sinh^2 ku}
	  			+ d\Omega^2\biggl]\biggr\}.
\ear
     For this metric, the allowed $\sigma$ range (\ref{6.4}) naturally
     splits into two parts: $\sigma < k$ and $\sigma \geq k$. One can
     easily show that for $\sigma < k$ particles moving along geodesics can
     arrive at the horizon $u=\infty$ in a finite proper time and may
     eventually (if geodesics can be extended) cross it, entering the BH
     interior  (type B1 BHs \cite{we}).  When, on the contrary, $\sigma\geq
     k$, the sphere $u=\infty$ turns out to be infinitely far and it takes
     an infinite proper time for a particle to reach it.  Since in the same
     limit $g_{22}\to \infty$, this configuration (a type B2 BH \cite{we})
     resembles a wormhole.

     In the case $[2-]$, $k = 0$, the BD metric containing a
     nonzero electric charge $q$ has the form
\bear                                                    \label{ds2-}
     ds^2 = \e^{2\sigma u}
     \biggl\{ \frac{h^2\,dt^2}{q^2\,\sinh^2[h(u+u_1)]} -
     			\frac{q^2 \,\sinh^2[h(u+u_1)]}
				{h^2 u^2}
          \biggr[\frac{du^2}{u^2}+ d\Omega^2\biggl]\biggr\}.
\ear
     The allowed range of the integration constants (\ref{6.6})
     again splits into two halves: for
     $\sigma < 0$ we deal with a type B1 BH, for $\sigma > 0$ with that of
     type B2 ($\sigma=0$ is excluded since it leads to $\phi=\const$, hence
     to GR).

     The properties of B1 and B2 structures are quite different and will be
     discussed separately.

\subsection{Type B1 \bhs: analytical extensions}

    Consider type B1 BHs for $k>0$. To obtain a Kruskal-like extension,
    we introduce, as usual, the null coordinates $v$ and $w$:
\beq
    v = t + x, \qquad w = t - x, \qquad
                    x \eqdef -\int \frac{\shhsq}{\sinh^2 ku}du  \label{vw}
\eeq
    where $x \to \infty$ as $u \to 0$ and $x \to -\infty$
    as $u \to \infty$. The asymptotic behaviour of
    $x$ as $u \to \infty$ is $x \sim \e^{2(h-k)u}$, and in a finite
    neighbourhood of the horizon $u=\infty$ one can write
\beq                                                        \label{x}
    x \equiv \half(v-w) = - \frac{k^2 q^2 \e^{2hu_1}}{2h^2 (h-k)}
	     \e^{2(h-k)u}\cdot f(u)
\eeq
    where $f(u)$ is an analytic function of $u$, with $f(\infty)=1$.

    To regularize the metric at the horizon,
    let us define new null coordinates $V<0$ and $W>0$
    related to $v$ and $w$ by
\beq                                                        \label{VW}
    - v = (-V)^{-n-1}, \quad\  w = W^{-n-1}, \quad\  n=\const.
\eeq
    The mixed coordinate patch $(V,w)$ is defined for $v<0$ ($t<-x$) and
    covers the whole past horizon $v=-\infty$. Similarly, the patch $(v,W)$
    is defined for $w>0$ ($t>x$) and covers the whole future horizon
    $w=+\infty$. So these patches can be used to extend the metric through
    any of the horizons.

    As is easily verified, a finite value of the metric coefficient $g_{vW}$
    at the future horizon $W=0$ is achieved if we take
\beq
    n+1 = (h-k)/(k-\sigma),                             \label{n}
\eeq
    which is positive for  $h>k>\sigma$. This provides as well a finite
    value of $g_{Vw}$ at the past horizon $V=0$.

    (If we tried to do the same for the B2 structure, we would find that
    the regularization is only achieved at $W=\infty$ and there are
    no values of $W$ where to continue the manifold.)

    One can now study the conditions of crossing, say, $W=0$
    from positive to negative values of $W$, since $W$ is an admissible
    coordinate on the future horizon. This coordinate is, however, defined
    explicitly only in the close neighbourhood of the horizon.
    To study the geometry in a finite or infinite region beyond the
    horizon, it is helpful to introduce a new radial coordinate $\rho(u)$
    behaving like $W$ near $W=0$. Indeed, let us introduce
    the coordinate $\rho$ for the metric (\ref{ds1-}) by
\beq                                                         \label{rho}
    \e^{-2ku} = \rho^{m-n}
\eeq
    where
\beq
    m = (h-k+\sigma)/(k-\sigma).                           \label{def-m}
\eeq
    As a result, the solution (\ref{ds1-}), defined originally in the
    static region ($\rho>0$), takes the form
\bear                                                \label{global}
    ds^2 \eql
    \frac{h^2}{q^2P_1^2(\rho)}\rho^{n+2}dt^2
    	-4q^2
    \biggl(\frac{m-n}{m+1}\biggr)^2 \frac{P_1^2(\rho)}{P_2^2(\rho)}
    	\Biggl[
    \frac{(m-n)^2}{P_2^2(\rho)}\rho^{-n-2}d\rho^2
				+ \rho^{-m}d\Omega^2\Biggr], \nn
    \phi \eql \rho^{m-n-1}, \\
    P_1(\rho) \al\eqdef\al \half \e^{hu_1}
    	       		\biggr[1 - \e^{-2hu_1}\rho^{m+1}\biggl],
    \cm    	P_2(\rho) \eqdef  1 - \rho^{m-n} .
\earn
    Due to (\ref{x}), $\rho$ is related to the mixed null coordinates
    $(v,W)$ by
\bearr                                                      \label{rhoW}
    \rho (v,W) = W\,\Bigl(1-vW^{n+1}\Bigr)^{-1/(n+1)}
		 (2f)^{1/(n+1)}
\ear
    This relation and a similar one giving $\rho (V,w)$
    show that when the future (past) horizon is crossed, $\rho$ varies
    smoothly, behaving like and $W$ or $V$ and changing its sign
    simultaneously with them.  For $\rho < 0$ the metric (\ref{global})
    describes the space-time regions beyond the horizons if the latter are
    regular.

    However, the metric (\ref{global}) makes sense at $\rho <0$
    only if the numbers $m$ and $n$ are both integers
    since otherwise fractional powers of negative numbers violate the
    analyticity as soon as the horizon is crossed.  This leads
    to a discrete set of ratios of the integration constants
    $h/k$ and $\sigma/k$:
\beq
     \frac hk = \frac{m+1}{m-n}, \cm  \frac{\sigma}k= \frac{m-n-1}{m-n}.
      						\label{qu}
\eeq
    where, according to the regularity conditions (\ref{6.4}),
    $m > n \geq 0$. Excluding the case $m=n+1$ that leads to $\sigma =0$,
    we see that BD BHs with regular horizons
    correspond to integers $m$ and $n$ such that
\beq
    m-2 \ge n \ge 0.                                          \label{mn}
\eeq

    We conclude that, although the curvature scalars
    are regular on the Killing horizon $u=\infty$, the metric cannot
    be extended beyond it unless the ratios $h/k$ and $\sigma/k$
    obey the ``quantization condition'' (\ref{qu}), and is generically
    singular. The Killing horizon, which is at a finite affine distance, is
    part of the boundary of the space-time, where geodesics and other
    possible trajectories terminate. Similar properties
    have been obtained in our previous papers \cite{kg1, kg2} for the
    vacuum case and earlier in a (2+1)--dimensional model with exact
    power--law metric functions \cite{sigma} and in the case of black
    $p$--branes \cite{GHT}.

    We have obtained a discrete family of BH solutions whose parameters
    depend on the two integers $m$ and $n$. This does not mean, however,
    that the observable parameters of the solution, the mass and the
    electric and scalar charges are ``quantized". Indeed, the electric
    charge remains to be an independent integration constant, the scalar
    field $\phi$ is well characterized by the the constant $\sigma$ given in
    (\ref{qu}) as a multiple of $k$, and the mass $M$ [cf. (\ref{GM})]
    in the present case reads:
\beq
    GM = \sqrt{h^2 + q^2} - \sigma                          \label{M1-}
	  = \sqrt{k^2 + q^2 + \sigma^2 |2\omega +3|} - \sigma.
\eeq
    Thus two constants, $k>0$, specifying the length scale of the solution,
    and the charge $q\ne 0$, remain arbitrary, and other constants are
    expressed in terms of them and the integers $m$ and $n$.  On the other
    hand, the coupling constant $\omega$ takes, according to (\ref{r2}) and
    (\ref{phi-BD}), discrete values:
\beq
	|2\omega+3| = \frac{(2m-n+1)(n+1)}{(m-n-1)^2}.          \label{om-mn}
\eeq

    Notably the \RN\ solution cannot be obtained from the present discrete
    family as a special case. Indeed, putting $m=n+1$, we obtain $\sigma=0$
    and $\omega=\infty$, that is, we abandon the BD theory; however, as
    follows from (\ref{om-mn}) and (\ref{phi-BD}), an expression for the
    scalar charge $C$ corresponding to the Einstein-frame field $\varphi$,
    remains finite: $C^2 = 2k^2 (n+1)(n+3)$, and we have still $h> k$,
    whereas for the \RN\ solution one must have $h=k$. We thus arrive at a
    subfamily of solutions of GR with an electric field and a minimally
    coupled scalar field. The \RN\ solution can be only recovered if we
    admit $n=-1$ and $m=0$.

The solution $[2-]$ ($k=0$) of the BD theory also has a Killing
horizon ($u \to \infty$) at a finite geodesic distance provided $\sigma<0$.
However, this space-time, as well as its vacuum counterpart, does not admit
a Kruskal--like extension and is therefore singular. The reason is that in
this case the relation giving the tortoise--like coordinate $x$,
\beq                                                          \label{x0}
   x = -\int \frac{q^2}{h^2} \frac{\shhsq}{u^2} du
	= -\frac{q^2 \e^{2hu_1}}{4h^2} \frac{\e^{2hu}}{u^2} [1+o(1)]
\eeq
(where $o(1)$ corresponds to the asymptotic $u\to\infty$)
cannot be used to obtain $u$ as an analytic function of $x$ near $u=\infty$.
The same happens to the coordinate $\rho$ which might be introduced in the
above manner to describe the region beyond the horizon since here, as
$u\to\infty$, \ $\rho = \const\times u^{-2}\e^{2\sigma u}[1+o(1)]$.

\subsection{Type B1 \bhs: causal structures}

    Let us return to the case $[1-]$. One can notice
    that by (\ref{x}) the radial coordinate $x$ is related to $\rho$ by
\beq
	x = - (\rho f)^{-n-1},                      \label{x-rho}
\eeq
    so that for odd $n$ the horizon as seen from region II ($\rho <0$) also
    corresponds to $x\to -\infty$. For even $n$, in region II $\rho$
    is also negative but remains to be a spatial coordinate, while the
    horizon corresponds to $x\to\infty$.  These observations are helpful in
    constructing the Penrose diagrams.

    The resulting causal structures depend on the parities of $m$ and $n$.

\medskip\noi
    {\bf [1--a]}. Both $m$ and $n$ are even, so $P_2(\rho)$ is an even
    function vanishing at $\rho=\pm 1$, where $\rho=+1$ is the ``old"
    spatial infinity and $\rho=-1$ is a new one. The only feature that
    makes the two regions $\rho > 0$ and $\rho < 0$ different is
    the function $P_1(\rho)$ which is everywhere regular and finite.
    The resulting Penrose diagram is similar to that for
    the extreme Kerr space-time, an infinite tower of alternating regions I
    and II (\fig 2).  All points of the diagram, except the boundary and
    the horizons, correspond to usual 2-spheres.

\medskip\noi
    {\bf [1--b]}. Both $m$ and $n$ are odd; then both $P_1(\rho)$ and
    $P_2(\rho)$ are even functions and $\rho$ ranges from $+1$ to $-1$. The
    regions I and II are now anti-isometric ($g_{\mu\nu}(-\rho) =
    -g_{\mu\nu}(\rho)$); the metric tensor in region II ($\rho < 0$) has
    the signature ($-++\,+$) instead of ($+--\,-$) in region I.
    Nevertheless, the Lorentzian nature of the space-time is preserved. The
    Penrose diagram is shown in \fig 3.
    The apparently acausal behaviour of geodesics like E1 can be avoided by
    assuming helicoidal space-time extension, see \cite{kg2}.

    In both cases $[1-a,b]$ the maximally extended space-times are
    globally regular%
\footnote{A globally regular extension of an extreme
    dilatonic black hole, with the same Penrose diagram as in our case
    [1--a], was discussed in \cite{GHT}.}.

 \Picture{60}
 {\cm\inch\ \
\unitlength=0.37mm
\special{em:linewidth 0.4pt}
\linethickness{0.4pt}
\begin{picture}(114.00,150.00)
\put(80.00,20.00){\line(1,1){30.00}}
\put(110.00,50.00){\line(-1,1){60.00}}
\put(50.00,110.00){\line(1,1){30.00}}
\put(80.00,140.00){\line(1,-1){30.00}}
\put(110.00,110.00){\line(-1,-1){60.00}}
\put(50.00,50.00){\line(1,-1){40.00}}
\put(80.00,20.00){\line(-1,-1){10.00}}
\put(30.00,10.00){\line(-1,1){10.00}}
\put(20.00,20.00){\line(1,1){30.00}}
\put(50.00,50.00){\line(-1,1){30.00}}
\put(20.00,80.00){\line(1,1){30.00}}
\put(50.00,110.00){\line(-1,1){30.00}}
\put(20.00,140.00){\line(1,1){10.00}}
\put(70.00,150.00){\line(1,-1){10.00}}
\put(80.00,140.00){\line(1,1){10.00}}
\put(101.00,110.00){\makebox(0,0)[cc]{I}}
\put(101.00,50.00){\makebox(0,0)[cc]{I}}
\bezier{152}(98.00,62.00)(83.00,50.00)(98.00,38.00)
\bezier{140}(62.00,38.00)(50.00,25.00)(59.00,10.00)
\bezier{144}(62.00,38.00)(75.00,52.00)(62.00,62.00)
\bezier{200}(62.00,62.00)(45.00,80.00)(62.00,98.00)
\bezier{136}(62.00,98.00)(74.00,111.00)(62.00,122.00)
\bezier{160}(62.00,122.00)(47.00,139.00)(60.00,150.00)
\bezier{252}(80.00,20.00)(90.00,56.00)(80.00,80.00)
\put(69.00,118.00){\makebox(0,0)[lc]{E1}}
\put(105.00,79.00){\makebox(0,0)[rc]{E2}}
\put(105.00,72.00){\makebox(0,0)[rc]{E3}}
\put(114.00,57.00){\makebox(0,0)[rb]{E5}}
\bezier{208}(80.00,20.00)(78.00,57.00)(67.00,67.00)
\bezier{200}(67.00,67.00)(53.00,79.00)(50.00,110.00)
\put(30.00,104.00){\makebox(0,0)[rc]{E4}}
\put(31.00,140.00){\makebox(0,0)[cc]{II}}
\put(31.00,80.00){\makebox(0,0)[cc]{II}}
\put(31.00,20.00){\makebox(0,0)[cc]{II}}
\put(33.00,101.00){\line(6,-1){19.00}}
\put(84.00,66.00){\line(2,1){12.00}}
\put(95.00,59.00){\line(1,0){10.00}}
\bezier{96}(58.00,58.00)(67.00,51.00)(58.00,42.00)
\bezier{260}(58.00,58.00)(35.00,80.00)(50.00,110.00)
\bezier{176}(58.00,42.00)(38.00,22.00)(46.00,8.00)
\put(30.00,55.00){\makebox(0,0)[cc]{E6}}
\put(36.00,56.00){\line(2,1){16.00}}
\end{picture}
}
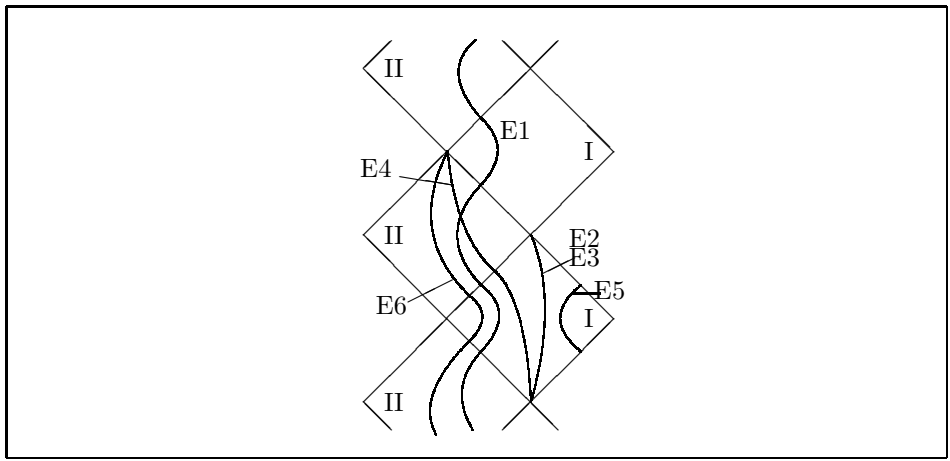
 {The causal structure of a BH with $m$ and $n$ both even.
 The curves E1--E6 depict various geodesics possible with this metric.}

\medskip\noi
    {\bf [1--c]}: $m$ even, $n$ odd. In region II ($\rho<0$)
    both $P_1$ and $P_2$ are positive, and the metric is regular up
    to $\rho\to -\infty$. In this limit the metric functions behave as
    follows:
\beq                                                       \label{lim-c}
    g_{tt}\sim |\rho|^{n-2m}, \cm
    			    g_{\rho\rho}\sim |\rho|^{-2m+3n},\cm
    g_{\theta\theta} \sim |\rho|^{2n+2-m}.
\eeq
    Evidently, as $\rho\to -\infty$, in all cases $g_{tt}\to 0$, while
    the area function $g_{\theta\theta}$ tends either to zero (for
    $m>2n+2$), or to a finite limit ($m=2n+2$), or to infinity ($m<2n+2$).
    In the first case this is a central singularity like Schwarzschilds's.
    If $m\leq 2n+2$, we can suspect one more new horizon and apply the
    above methodology to study its nature and to try to cross it.

    In short, one can introduce one more radial coordinate $\eta(\rho)$
    which behaves in the same way as a null coordinate providing a finite
    metric coefficient at a possible horizon. Such a coordinate can be
    determined from the asymptotic condition
    $|g_{tt}|\sim |g_{\eta\eta}|^{-1}$ as $\rho\to -\infty$.
    This is achieved by substituting
\beq
	\rho = -\eta^{-1/p}, \cm p= 2m-2n-1 > 0,           \label{eta}
\eeq
    the limit $\rho\to -\infty$ corresponds to $\eta\to +0$, and a
    transition across a possible horizon should be described by passing
    from positive to negative $\eta$. In terms of $\eta$ the
    metric coefficients behave in the following way:
\bear
	g_{tt} \sim (g_{\eta\eta})^{-1}
	                 \al\sim\al  \eta^{a_1},
	 			\cm    a_1 = 1 + (n+1)/p,     \nnv
	g_{\theta\theta} \al\sim\al  \eta^{a_2},
	 			\cm    a_2 = 1 - (m+1)/p.
\ear
    Just as before, a transition to negative $\eta$ makes sense only if both
    exponents $a_1$ and $a_2$ are integers. One can, however, notice that
    the sum
\beq                                                        \label{fraction}
	a_1 + a_2 = 2 - \frac{m-n}{p} = \frac{3}{2}- \frac{1}{2p}
\eeq
    is non-integer as long as $p= 2m-2n-1 > 1$. For the relevant $m$ and $n$
    the number $p$ can be equal to  5, 9, 13,...
    We conclude that in all nontrivial cases at least one of the exponents
    $a_1$ and $a_2$ is a fraction and hence there is no analytic
    extension beyond $\rho=-\infty$.

    The Kretschmann scalar tends to a finite limit as $\rho\to -\infty$
    if $m \leq \frac{3}{2}n + 1$, otherwise it diverges. However, even for
    $m$ and $n$ such that it is finite, the space-time terminates due
    to analyticity violation.

    In all cases $[1-c]$ the singularity at $\rho=-\infty$ is null,
    therefore the Penrose diagram does not repeat that for the
    Schwarzschild metric, but, instead, coincides with that of \fig 3, where
    now the outer boundaries of regions II depict singularities.

    It is of interest that the \RN\ solution, described in this scheme
    by the values $m=0,\ n=-1$ outside the allowed range (\ref{mn}), also
    belongs to class $[1-c]$; in this case the surface $\rho=-\infty$ is
    regular and corresponds to the well-known Cauchy horizon of the \RN\
    metric, and a further continuation proceeds as described in the
    textbooks.

\medskip\noi
    {\bf [1--d]}: $m$ odd, $n$ even.
    In region II the range of $\rho$ terminates at a zero of the
    function $P_1(\rho)$ where one finds a \RN-like repulsive
    ($g_{tt}\to\infty$) central ($g_{\theta\theta}\to 0$) singularity.  The
    resulting Penrose diagram is similar to that of the extreme ($q^2 =
    GM^2$) \RN\ space-time, with the difference that now the 4-dimensional
    metric changes its signature when crossing the horizon, similarly to
    case $[1-b]$, therefore the singularity should be interpreted
    as a spacelike one.

 \Picture{60}
 {\inch\ \ \
\unitlength=0.37mm
\special{em:linewidth 0.4pt}
\linethickness{0.4pt}
\begin{picture}(140.00,140.00)
\put(80.00,20.00){\line(1,1){30.00}}
\put(110.00,50.00){\line(-1,1){60.00}}
\put(50.00,110.00){\line(1,1){30.00}}
\put(80.00,140.00){\line(1,-1){30.00}}
\put(110.00,110.00){\line(-1,-1){60.00}}
\put(50.00,50.00){\line(-1,1){30.00}}
\put(20.00,80.00){\line(1,1){30.00}}
\put(50.00,50.00){\line(1,-1){30.00}}
\put(80.00,20.00){\line(0,0){0.00}}
\put(110.00,50.00){\line(1,1){30.00}}
\put(140.00,80.00){\line(-1,1){30.00}}
\bezier{256}(110.00,110.00)(121.00,71.00)(110.00,50.00)
\bezier{124}(129.00,69.00)(118.00,80.00)(129.00,91.00)
\bezier{124}(90.00,90.00)(80.00,102.00)(70.00,90.00)
\bezier{112}(70.00,90.00)(60.00,80.00)(70.00,70.00)
\bezier{124}(70.00,70.00)(80.00,58.00)(90.00,70.00)
\bezier{120}(90.00,70.00)(101.00,80.00)(90.00,90.00)
\bezier{208}(96.00,96.00)(110.00,81.00)(110.00,50.00)
\bezier{196}(96.00,96.00)(83.00,110.00)(80.00,140.00)
\bezier{132}(80.00,139.00)(77.00,122.00)(80.00,107.00)
\bezier{136}(80.00,140.00)(82.00,117.00)(80.00,106.00)
\put(80.00,61.00){\makebox(0,0)[ct]{E1}}
\put(130.00,110.00){\makebox(0,0)[cb]{E2}}
\put(130.00,106.00){\makebox(0,0)[ct]{E3}}
\put(136.00,91.00){\makebox(0,0)[lb]{E5}}
\put(99.00,129.00){\makebox(0,0)[lb]{E4}}
\put(58.00,134.00){\makebox(0,0)[rb]{E7}}
\put(61.00,132.00){\line(6,-1){18.00}}
\put(97.00,127.00){\line(-5,-2){14.00}}
\put(125.00,107.00){\line(-5,-2){13.00}}
\put(125.00,86.00){\line(5,2){9.00}}
\bezier{184}(80.00,140.00)(70.00,109.00)(61.00,99.00)
\bezier{204}(61.00,99.00)(44.00,80.00)(61.00,61.00)
\bezier{196}(61.00,61.00)(80.00,43.00)(80.00,20.00)
\put(45.00,124.00){\makebox(0,0)[cc]{E6}}
\put(49.00,122.00){\line(6,-1){23.00}}
\put(30.00,80.00){\makebox(0,0)[cc]{I}}
\put(132.00,80.00){\makebox(0,0)[cc]{I}}
\put(100.00,110.00){\makebox(0,0)[cc]{II}}
\put(100.00,50.00){\makebox(0,0)[cc]{II}}
\end{picture}
}
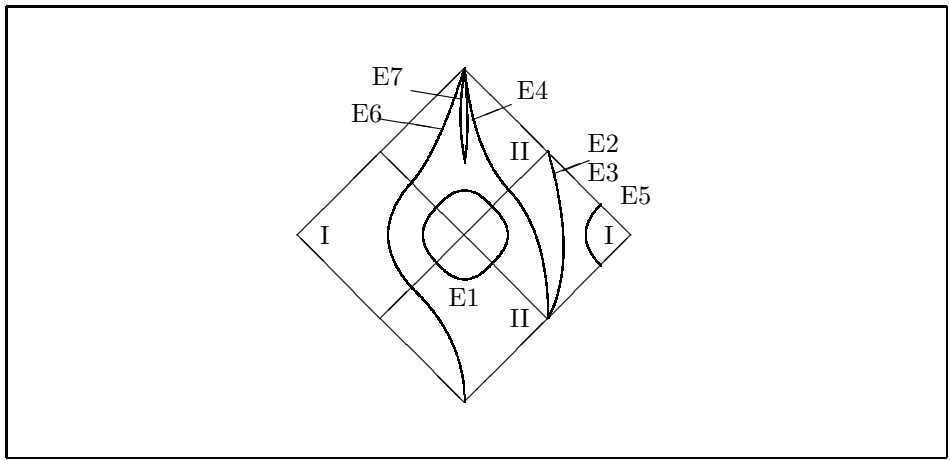
 {The causal structure of a BH with $m$ and $n$ both odd.
 The curves E1--E7 depict various geodesics possible with this metric.}

\subsection{Type B2 structures}

    For $k>0$, a type B2 structure occurs when $h>\sigma>k$. As before,
    the metric is transformed according to (\ref{vw})--(\ref{VW}) and at the
    future null limit (now infinity rather than a horizon, so we
    avoid the term ``\bh")  where now $W\to \infty$
    the asymptotic form of the metric is
\beq
	ds^2 =  -C_1 dv\,dW - C_2 W^{-m} d\Omega^2      \label{B2vW}
\eeq
    where $C_{1,2}$ are some positive constants, while
    the constant $m$, defined in (\ref{def-m}), is
    now negative.  A further application of the $v$-transformation
    (\ref{VW}) at the same asymptotic, valid for any finite $v<0$, leads to
\bearr                                                     \label{B2VW}
	ds^2 = -C_1 (-V)^{(h-\sigma)/(\sigma-k)}dV\,dW - C_2 W^{-m}d\Omega^2.
\ear
    If we now introduce new radial ($R$) and time ($T$) coordinates
    by $T=V+W$ and $R=V-W$, in a spacelike section $T=\const$ the limit
    $R\to -\infty$ corresponds to simultaneously $V\to -\infty$ and
    $W\to +\infty$, with $|V|\sim W$, and the metric (\ref{B2VW}) turns into
\bear
	ds^2 \eql 4C_1 (-R)^{(h-\sigma)/(\sigma-k)}(dT^2 - dR^2) \label{B2RT}
                           - C_2 (-R)^{-m} d\Omega^2.
\ear
    This asymptotic is a nonflat spatial infinity, with infinitely
    growing coordinate spheres and also $g_{00}\to \infty$, i.e.,
    this infinity repels test particles.

    A Penrose diagram of a B2 type configuration coincides with a single
    region I in any of the above diagrams; all its sides depict null
    infinities, its right corner corresponds to the usual spatial infinity
    and its left corner to the unusual one, represented by the metric
    (\ref{B2RT}).

    A similar picture is obtained for type B2 structures in the case $[2-]$
    ($k=0,\ \sigma > 0$).

\subsection{Comparison with Brans-Dicke vacuum}

    Vacuum BD configurations are easily obtained from the charged ones in the
    limit $q\to 0$, as outlined at the end of \sect 3. The analysis of the
    case $[1-]$ was a little more transparent in Refs.\,\cite{we,kg1,kg2}
    because for $k>0$, after the coordinate transformation
\beq
     \e^{-2ku} = 1 - 2k/r \equiv P(r)                       \label{def-P}
\eeq
     the solution took the form
\bear                                                       \label{ds+}
     ds^2 \eql P^{-\xi}\Bigl(P^{a }dt^2 - P^{-a }dr^2
                                 - P^{1 - a }r^2 d\Omega^2 \Bigl),\nn
     \phi \eql P^\xi
\ear
     with the constants related by
\beq
     (2\omega+3) \xi^2 = 1-a^2, \cm a=h/k, \cm  \xi = \sigma/k.
							    \label{int+}
\eeq
    and the further consideration was conducted in terms of $P$;
    the horizon took place at $P=0$. One can easily verify that our
    present conditions for the occurence of type B1 and B2 structures are
    still valid for the vacuum case and reduce to those of \cite{kg1, kg2}
    with the notations (\ref{int+}). The same applies to the parities of
    $m$ and $n$ in the classification $[1-a]$--$[1-d]$. There are only some
    differences in the description of particular cases. Thus, in the vacuum
    case $[1-a]$ not only the qualitative behaviour of the solution is
    symmetric with respect to $\rho=0$, but even the transition $\rho\to
    -\rho$ is an isometry. A description of the $[1-b]$ case is
    unchanged. For $[1-c]$ the vacuum solution behaves simpler: at
    $\rho=-\infty$ there is always a central spacelike singularity and the
    Penrose diagram repeats that for the Schwarzschild metric. Lastly, for
    $[1-d]$, the vacuum solution is singular at $\rho=-\infty$ (as for
    $[1-c]$ and unlike charged $[1-d]$) and the singularity is again central
    and spacelike; the Penrose diagram coincides with that of charged
    $[1-d]$, and there is the same signature change when crossing the
    horizon.

\section{Stability}

    A study of small (linear) \sph \pns\ of the above static solutions (or
    static regions of the charged BHs) is to a large extent similar to
    that of vacuum systems described in \cite{kg1, kg2}, therefore we here
    omit some details of the method but give the results completely.

    We now consider, instead of $\varphi(u)$, a perturbed unknown function
\[
	\varphi(u,t)= \varphi(u)+ \df(u,t)
\]
    and similarly for the metric functions $\alpha,\beta,\gamma$, where
    $\varphi (u)$, etc., are taken from the static solutions of \sect 2.
    The electromagnetic field is, by assumption, still governed by the
    potential component $A_t$, therefore it does not invoke a new dynamical
    degree of freedom as compared with the vacuum case. We are working in
    the Einstein conformal frame. The consideration applies to the whole
    class of STT (\ref{L1}); its different members can differ only in
    boundary conditions to be satisfied by the \pns.

    We use the gauge freedom existing in the perturbation analysis
    (a choice of the frame of reference and the coordinates in
    the perturbed space-time) by putting
\beq
	\da = 2\db + \dg,                                     \label{da}
\eeq
    thus extending to \pns\ the harmonic coordinate condition of the static
    system in the Einstein conformal frame. In this and only in this case
    the scalar equation due to (\ref{L2}) for $\df$ ($\DAL \varphi =0$)
    decouples from the other \pn\ equations and reads
\beq
	\e^{4\beta} \delta\ddot\varphi  - \df''=0.          \label{edf}
\eeq
    Here $\e^{2\beta} = - g^{\rm E}_{\theta\theta}$ is the area function of
    the unperturbed solution in the Einstein frame, dots denote $d/dt$ and
    primes, as before, $d/du$. Since the scalar field is the only dynamical
    degree of freedom, \eq(\ref{edf}) can be used as the master one, while
    other equations due to (\ref{L2}) only express the metric and
    electromagnetic variables in terms of $\df$, provided the whole set of
    field equations is consistent.  That it is indeed the case, is directly
    verified.

    The static nature of the background solution makes it possible to
    separate the variables in \eq (\ref{edf}),
\beq
	\df = \psi(u) \e^{i\omega t},                    \label{psi}
\eeq
    and to reduce the stability problem to a boundary-value problem for
    $\psi(u)$. Namely, if there exists a nontrivial solution to (\ref{edf})
    with $\omega^2 <0$, satisfying some physically reasonable conditions at
    the ends of the range of $u$, then the static system is unstable since
    $\df$ can exponentially grow with $t$.  Otherwise it is stable in the
    linear approximation.

    Suppose $-\omega^2 = \Omega^2,\ \Omega > 0$. In what follows we
    use two forms of the radial equation (\ref{edf}): the one directly
    following from (\ref{psi}),
\beq
	\psi'' -\Omega^2 \e^{4\beta(u)}\psi=0,             \label{epsi}
\eeq
    and the normal Liouville (Schr\"odinger-like) form
\bearr
	d^2 y/dx^2 - [\Omega^2+V(x)] y(x) =0,   \nnn \inch
	V(x) = \e^{-4\beta}(\beta''-\beta'{}^2).            \label{ey}
\ear
    obtained from (\ref{epsi}) by the transformation
\beq
	\psi(u) = y(x)\e^{-\beta},\qquad                     \label{tx}
				x = - \int \e^{2\beta(u)}du.
\eeq
    Here, as before, a prime denotes $\d/\d u$. It is of interest to note
    that $x$ is the same ``tortoise" coordinate that was used for continuing
    the \bh\ metrics through horizons, see \eq (\ref{vw}).

    The boundary condition at spatial infinity ($u\to 0$, $x \simeq 1/u \to
    +\infty$) is evident: $\df\to 0$, or $\psi\to 0$.
    For our metric (\ref{s2}) the effective potential $V(x)$ has the
    asymptotic form
\beq
    V(x) \approx 2h/x^3, \cm {\rm as} \cm x\to +\infty,
\eeq
    hence the general solutions to (\ref{ey}) and (\ref{epsi}) have the
    asymptotic form
\bear
    y \al\sim \al c_1\e^{\Omega x} + c_2\e^{-\Omega x}
    \qquad (x \to +\infty),                             \label{as+} \\
    \psi \al                                                \label{aspsy}
        \sim \al u \bigl(c_1 \e^{\Omega/u} + c_2\e^{-\Omega/u}\bigr)
    \qquad (u \to 0),
\ear
    with arbitrary constants $c_1,\ c_2$. Our boundary condition leads to
    $c_1=0$.

    For $u\to \umx$, where in many cases the background field $\varphi$
    tends to infinity, the boundary condition is not so evident.
    Refs.\,\cite{hod,bm} and others, dealing with minimally coupled or
    dilatonic scalar fields, used the minimal requirement providing the
    validity of the \pn\ scheme in the Einstein frame:
\beq
    |\df/\varphi| < \infty.                                \label{weak}
\eeq
    In STT, where Jordan-frame and Einstein-frame metrics are related by
    $g^{\rm J}_{\mu\nu} = (1/\phi)g^{\rm E}_{\mu\nu}$,
    it seems reasonable to require that the perturbed conformal factor
    $1/\phi$ behave no worse than the unperturbed one, i.e.
\beq
    |\delta\phi/\phi| < \infty.                         \label{strong}
\eeq
    An explicit form of this requirement depends on the specific STT and
    can differ from (\ref{weak}), for example, in the BD theory, where
    $\phi$ and $\varphi=Cu$ are connected by (\ref{phi-BD}), the requirement
    (\ref{strong}) leads to $|\df| <\infty$. We will refer to (\ref{weak})
    and (\ref{strong}) as to the ``{\it weak\/}" and ``{\it strong\/}"
    boundary condition, respectively. For systems where both $\phi$
    and $\varphi$ are regular at $u\to \umx$ these conditions coincide
    and both give $|\df|<\infty$.

    Let us now discuss different cases of the STT solutions under study.
    We will suppose that the scalar field $\phi$ is regular for
    $0<u<\umx$, so that the conformal factor $\phi^{-1}$ in (\ref{s2})
    does not affect the range of the $u$ coordinate.

\medskip\noi
    {\bf [1\,+]},
    the singular solution of normal STT. As $u \to +\infty$,
    $\beta \sim (h-k)u \to -\infty$, so that $x$ tends to a finite limit and
    it is convenient to suppose $x\to 0$. The effective potential $V(x)$ then
    behaves as $V \sim -1/(4x^2)$, and the general asymptotic solution to
    (\ref{ey}) leads to
\beq
    \psi(u) \approx y(x)/\sqrt{x} \approx (c_3 + c_4\ln x) \quad
    (x \to 0).             \label{y1}
\eeq

    The weak boundary condition leads to the requirement
    $|\delta\varphi/\varphi| \approx |y|/(\sqrt{x}|\ln x|) < \infty$, met by
    the general solution (\ref{y1}) and consequently by its special solution
    that joins the allowed case ($c_1=0$) of the solution (\ref{as+}) at the
    spatial asymptotic. We then conclude that the static field configuration
    is unstable, in agreement with the previous work \cite{hod}.

    As for the strong boundary condition (\ref{strong}),
    probably more appropriate in STT, its explicit form
    varies from theory to theory, and a general conclusion is impossible.
    In the special case of the BD theory the condition (\ref{strong})
    means $|\psi|< \infty$ as $u\to +\infty$. Such an asymptotic behavior
    is forbidden by \eq (\ref{epsi}), according to which $\psi''/\psi > 0$,
    i.e. the function $\psi(u)$ is convex and so cannot be bounded as $u
    \to \infty$ for an initial value $\psi(0) = 0$ ($c_1 = 0$). We
    conclude that the BD static system is stable.

    Thus in this singular case the choice of a boundary condition is crucial
    for the stability conclusion. In GR with a minimally coupled scalar
    field \cite{hod} there is no reason to ``strengthen" the weak condition
    that leads to the instability. In the BD case the strong condition seems
    more reasonable and implies stability. For any other STT the situation
    must be considered separately.

\medskip\noi
    {\bf [2\,+]}.
    With slightly more effort, the results of item $[1+]$ are reproduced,
    and a stability conclusion again depends on the boundary conditions.

\medskip\noi
    {\bf [3\,+]},
    the case of \RN-like central singularities in normal theories.
    Here $\umx < \infty$, and we assume that $|\phi(\umx)| < \infty$.
    (We exclude possible pathological cases
    of zero or infinite $\phi$ at $u=\umx<\infty$ which can
    be considered specially if necessary.) Then the weak and strong
    conditions concide.  As $u\to\umx$, we can put again $x\to 0$ and it
    appears then that $V(x) \simeq -2/(9x^2)$. The general solution to \eq
    (\ref{ey}) behaves as
\beq
	y = c_1 x^{1/3} + c_2 x^{2/3}
\eeq
    near $x=0$, whereas the boundary condition is $yx^{-1/3} < \infty$.
    Since this condition is satisfied  for any constants $c_1,\ c_2$, we
    conclude that (generically) this type of solution is unstable in any
    STT.

\medskip\noi
    {\bf [1\,--], [2\,--]}.
    This case includes singular solutions and cold \bhs\ as exemplified
    above for the BD theory.

    As $u\to +\infty$, $\beta \to +\infty$, so that $x\to
    -\infty$ and $V(x) \to 0$. The general solution to \eq
    (\ref{ey}) again has the asymptotic form (\ref{as+}) for $x \to
    -\infty$.  The weak condition (\ref{weak}) leads, as in the previous
    case, to the requirement $|y|/(\sqrt{|x|}\ln|x|) <\infty$, and, applied
    to (\ref{as+}), to $c_2=0$. This means that the function $\psi$ must
    tend to zero for both $u\to 0$ and $u\to \infty$, which is impossible
    due to $\psi''/\psi >0$. Thus the static system is stable. Obviously the
    more restrictive strong condition (\ref{strong}) can only lead to the
    same conclusion.

\medskip\noi
    {\bf [3\,--]}.
    In the generic case the solution describes a
    wormhole, and in the exceptional case (\ref{4.6}) there is a cold
    \bh\ with a finite horizon area. In all such cases, as $u\to\umx =
    \pi/|k|$, one has $x\to -\infty$ and $V \sim 1/|x|^3 \to 0$, so that the
    stability is concluded just as in the cases $[1-]$, $[2-]$.

\medskip\noi
    {\bf [4\,--]}.
    The results differ for different cases a,b,c described in \sect 5 (and
    this description applies to all STT under our assumptions). Thus, in the
    singular case $[4-a]$ we repeat the instability conclusion made for
    $3+]$. In the \wh\ case $[4-b]$ we obtain stability just as for $[3-]$.
    Lastly, for the ``horn" $[4-c]$ we have a finite potential at $x=0$
    corresponding to $u=\umx$ and a finite general solution for $\psi(u)$,
    hence instability.

    In the vacuum case we are restricted to the above variants $[1+]$,
    $[1-]$, $[2-]$, $[4-b]$ with their corresponding stability conclusions.

\section{Concluding remarks}

    We can conclude by the following observations.
\begin{enumerate}
\item
    Black holes do exist in anomalous scalar-tensor theories,
    i.e., when the kinetic term of the scalar field is
    negative, contrary to what was sometimes claimed \cite{kim}.
\item
    For $k\geq 0$, there are no conventional (type A) BHs, but there exist
    BHs with an infinite area (type B), as confirmed explicitly for a
    special case --- the BD theory.  They in turn
    split into two classes, B1 and B2, with, respectively, finite and
    infinite proper time needed for an infalling particle to reach a
    horizon. Type B2 structures do not need an analytic extension and
    resemble \whs\ in that they possess another spatial asymptotic.
\item
    In the case $k<0$ type A BHs can exist, but only in theories with
    variable $\omega$, and such explicit examples are yet to be found.
\item
    Type 1 Brans-Dicke BHs generically possess singular horizons, the
    singularity being caused by analyticity violation. Only a descrete
    family of solutions, parametrized by two integers, $m$ and $n$,
    describes BHs with traversable horizons.
\item
    From the above relations one can observe that at the second asymptotic
    of all type B2 configurations the BD scalar field $\phi\to 0$, i.e., the
    effective gravitational coupling tends to infinity.
    The same happens at traversable horizons of type B1 BHs, and, moreover,
    the effective coupling $\phi^{-1}$ is negative in regions II when
    $m$ and $n$ have equal parity, i.e. in the cases $[1-a]$ and $[1-b]$.
    At singularities of B1 configurations with $\sigma<0$, on the contrary,
    $\phi\to \infty$ and the gravitational coupling vanishes.
\item
    The electric charge adds some kinds of solution behaviour as compared
    with the vacuum case but does not drastically change the situation with
    BHs.
\item
    Despite their exotic properties, the BH solutions found here are
    stable, at least with respect to small radial \pns.
\item
    For non-BH solutions in normal STT stability conclusions crucially
    depend on the boundary condition adopted for \pns\ at singularities. Old
    results on the instability of solutions with scalar fields in GR
    \cite{hod} are confirmed.
\item
    The Brans-Dicke BHs under consideration have infinite horizon areas
    and zero Hawking temperature. This suggest an infinite entropy,
    consistently with the fact that BHs have negative specific heat.
    However, a precise calculation of the entropy requires the
    determination of the surface term in the gravitational action.
    In the case of an STT, this surface term differs
    from the usual one by the presence of the scalar field,
    making the usual expression for the entropy inappropriate.
    Therefore such a calculation requires a separate study.
\item
    Tidal forces become infinite at horizons with infinite areas. Hence,
    only a point particle can cross such a horizon without being destroyed,
    just as in the vacuum case \cite{kg2}.

\end{enumerate}

\vspace{0.5cm}

\section*{Appendix. On horizon regularity conditions}
\renewcommand{\theequation}{A.\arabic{equation}}
\sequ 0

An event horizon is, by definition, a regular surface, which implies finite
values of all curvature invariants. The finiteness of the Kretschmann
scalar $R^{\mu\nu\lambda\gamma}R_{\mu\nu\lambda\gamma}$ is known to be the
most efficient criterion of regularity.

Using it, we will prove that (at least for static, \sph space-times) an
infinite Hawking temperature $\TH$ of an assumed horizon indicates that it
is a curvature singularity rather than a horizon (Lemma 1)%
    \footnote{Although Lemma 1 has been proved \cite{bim} in a more
              general $D$-dimensional setting, it seems useful to present
	      it here for $D=4$. Besides, the expressions for $K_i$ and
	      $\TH$ are used in the text of the paper.}.
Another simple result (Lemma 2) is that $\TH =\infty$ --- hence there is a
singularity --- if an assumed horizon is visible for a static observer,
i.e., the integral $t^* = \int \e^{\alpha - \gamma}du$ converges.

Thus Criteria C3 and C4 from \sect 2 are simple and convenient necessary
conditions of horizon regulatity.

     The Kretschamnn scalar for the metric (\ref{m1}) may be written as
\beq                                                        \label{k}
	K = 4K_1^2 + 8K_2^2 + 8K_3^2 + 4K_4^2
\eeq
     where
\bear                                                       \label{A2}
K_1 &=& {R^{01}}_{01} = - \e^{-\alpha - \gamma}
         \biggr(\gamma'\e^{\gamma - \alpha}\biggl)'  ,\nn
K_2 &=& {R^{02}}_{02} = {R^{03}}_{03} = - \e^{-2\alpha}\beta'\gamma' ,\nn
K_3 &=& {R^{12}}_{12} = {R^{13}}_{13} = - \e^{-\alpha -
\beta}\biggr(\beta'
         \e^{\beta-\alpha}\biggl)' \, , \nn
K_4 &=& {R^{23}}_{23} = \e^{-2\beta} - \e^{-2\alpha}{\beta'}^2
\ear
     where a prime denotes $d/du$.
     The structure of \eq(\ref{k}) indicates that an infinite value of any
     $K_i$ implies the presence of a singularity at a given point of the
     space-time.

     On the other hand, using e.g. formulae from the book \cite{Wald}, one
     finds for static metrics written in the form (\ref{m1}) the following
     expression for the Hawking temperature of a surface $u=u^*$ where
     $\e^{\gamma}=0$, assumed to be a horizon:
\beq
     \TH = \frac{\kappa^*}{2\pi}, \cm
     \kappa^* \eqdef \lim_{u\to u^*} \kappa(u), \cm
     \kappa(u) \eqdef \e^{\gamma-\alpha}|\gamma'|              \label{th}
\eeq
     where we have put the Boltzmann constant $k_{\rm B}$ and the Planck
     constant $\hbar$ equal to 1.  (The same expression can be obtained
     using other methods, such as Euclidean continuation of the metric).

     We are now ready to prove the following two lemmas.

\Theorem{Lemma 1}
{If, at a certain surface $u = u^* $ of a static, spherically symmetric
space-time with the metric (\ref{m1}), $\e^{\gamma}= 0$ (a candidate
horizon) and the Hawking temperature $\TH$ calculated for $u=u^*$, is
infinite, this surface is a curvature singularity.}

     By assumption, $\e^{\gamma} \to  0$ when $u \to u^*$.  Assume, in
     addition, that $\kappa^* =\infty$, while both functions $\gamma(u)$ and
     $\kappa(u)$ are monotonic in some neighbourhood of $u^*$.  Let us show
     that then the Kretschmann scalar $K \to  \infty$ as $u \to u^*$.

     It is sufficient to prove that $K_1 \to \infty$.

     Let us use the fact that the expressions $K_1$ (as well as other $K_i$)
     and $\kappa(u)$ are unaffected by reparametrizations of the radial
     coordinate $u$.  With this invariance, any coordinate conditon for $u$
     may be chosen without loss of generality. Let us choose the following
     one:
\beq
     \gamma + \alpha = 0 \, .         \label{A3}
\eeq
     Then
\[
     K_1 = - \Half [2\gamma'\e^{2\gamma}]' = - \Half [\e^{2\gamma}]''.
\]
     By our assumptions we have $\e^{2\gamma} \to  0$ and
     $(\e^{2\gamma})' \to  \infty$ as $u \to  u^*$.

     Let us denote $g(u) = \e^{2\gamma}$, ${1}/{g'(u)} = G(g)$.
     Then $G(g) \to  0$ as $g \to 0$.
     On the other hand, one can write:
\[
	\frac{dg}{du} = \frac{1}{G(g)} \quad\then \quad u = \int G(g)dg.
\]
     This integral is evidently finite, hence $u^*$ is finite in the
     coordinates (\ref{A3}). Thus, for a finite value of $u$, we have
     $g'= {dg}/{du} \to  \infty$, therefore
\[
	 g'' \to  \infty \quad  \then  \quad   |K_1| \to  \infty,
\]
     which proves Lemma 1.

\Theorem{Lemma 2}
{If, at a candidate horizon $u = u^*$ ($\e^{\gamma(u^*)}=0$) of a static,
spherically symmetric space-time with the metric (\ref{m1}), the integral
$t^* =\int \e^{\alpha - \gamma}du$ converges, then at $u = u^*$ the
temperature $\TH=\infty$.}

     Let us again use the coordinate freedom and put $\alpha\equiv \gamma$.
     Then we have simply
\[
     t^* = \int du, \cm   \kappa(u) = |\gamma'(u)|.
\]
     So the convergence of $t^*$ means just $|u^*| < \infty$, which is
     compatible with $\gamma(u^*)=-\infty$ only if $\gamma'(u^*) =\infty$,
     whence $\kappa^* = \infty$. Lemma 2 is proved.

\medskip\noi
{\bf Comment.}
The very notion of a horizon implies that it must be in absolute past or
future for an observer at rest in a static space-time, and, moreover,
it is physically clear that $\TH=\infty$ must mean that such a configuration
immediately evaporates and actually cannot exist.
These considerations, however, rest on physical interpretations, whereas
Lemmas 1 and 2 are of purely geometric nature and provide certain
mathematical grounds for such interpretations.

\bigskip\noi
{\bf Acknowledgements.} We would like to thank George Matsas
and Jos\'e P. Lemos for many helpful discussions. We are especially grateful
to G\'erard Cl\'ement who, in our collaboration on Refs.\,\cite{kg1, kg2},
expressed some ideas used as well in this paper. K.B. acknowledges the
kind hospitality of the colleagues from DF-UFES, Vit\'oria, Brazil,
during his stay while this paper was completed. The work
was supported in part by CNPq (Brazil) and CAPES (Brazil).

\end{document}